\documentclass[prd,showpacs,showkeys,secnumarabic,12pt]{revtex4}
\usepackage{amsmath,amsbsy}

\newcommand{\pni}{\par\noindent}
\begin{document}
\title{Time-dependent extra dimension and higher-dimensional 
modifications to the matter content in FRW spacetimes } 
\author {M. La Camera} \email{lacamera@ge.infn.it} \affiliation 
{Department of Physics and INFN - University of Genoa\\Via 
Dodecaneso 33, 16146 Genoa, Italy}
\begin{abstract} \pni
In this work we suggest that higher-dimensional  modifications  
to the matter content in FRW spacetimes can be obtained not only,
as first considered by Ponce de Leon, referring to ``moving''   
4D hypersurfaces non-orthogonal to the time-dependent extra 
dimension of an embedding 5D manifold, but also referring to  
``fixed'' 4D hypersurfaces orthogonal to a suitable scalar 
function which defines a static foliation of the 5D manifold and 
takes the role of the extra dimension in a suitable coordinate 
system. Results obtained in each approach  crucially  depend on 
the method used to identify the 4D metric of our brane universe 
from the 5D metric of the bulk manifold. 
\end{abstract} 
\maketitle  
\section{I\lowercase{ntroduction}} \pni 
Recently Ponce de Leon  ${}^{1,2,3}$  showed that our observable
universe can be devised as a dynamic four-dimensional 
hypersurface which depends  explicitly on the evolution of the 
time-dependent extra dimension of the embedding five-dimensional 
manifold and is non-orthogonal to it. As a consequence  it is 
possible to construct a four-dimensional model which predicts 
higher-dimensional modifications to the energy-momentum tensor as
obtained by the usual form of the Friedmann-Robertson-Walker 
(FRW) metric. However, since  there is more than one way for 
embedding a four-dimensional spacetime in a given 
five-dimensional manifold, the results obtained  crucially depend
on the method used to identify the 4D metric from the 5D one. In 
this paper we show that higher-dimensional modifications to  
``conventional'' FRW spacetimes can be obtained not only 
referring to ``moving'' hypersurfaces non-orthogonal to the extra
dimension but also considering ``fixed'' hypersurfaces orthogonal
to a suitable scalar function which defines a static covariant 
foliation of the bulk and takes the role of the extra dimension 
in a suitable coordinate system. We consider a five-dimensional  
manifold embedding a homogeneous and isotropic universe and 
described by a 5D metric with a time-dependent extra  dimension 
and utilize the geometric construction performed by Sehara and 
Wesson ${}^{4-6}$ to obtain the foliation of the manifold by 
static 4D hypersurfaces. Then we transform the previous 5D metric
into a metric where the new  extra coordinate does not 
depend on the new time and verify that also on each fixed
leaf of the foliation the induced 4D metric predicts 
higher-dimensional modifications with respect to the usual FRW 
line element. Finally we apply our approach to a well-known 
five-dimensional metric found by Ponce de Leon ${}^7$ and discuss
the results obtained in the different models. \\ \pni 
\textit{Conventions}. Throughout the paper the 5D metric 
signature is taken to be $(+,+,+,-,\varepsilon)$ where 
$\varepsilon$ can be $+1$ or $-1$ depending on whether the extra 
dimension is spacelike or timelike, while the choice of the 4D 
metric signature is $(+,+,+,-)$. Bulk indices will be denoted by 
capital Latin letters  and brane indices by lower Greek letters. 
Finally we use units where $c=G=1$. \pni
\section{F\lowercase{rom a time-dependent extra dimension to 
static 4\uppercase{D} hypersurfaces}}
\pni Our homogeneous  and isotropic universe is   envisaged as 
embedded in a five-dimensional manifold $M$ covered by an 
arbitrary system of coordinates $x^A=(r,\vartheta,\varphi,t,y)$. 
The 5D line element will be written in the usual form as 
\begin{equation} 
ds_5^2 = a^2(t,y)\,d\sigma_k^2 - n^2(t,y)\,dt^2 + 
\varepsilon\,b^2(t,y)\,dy^2 
\end{equation} 
where 
\begin{equation}
d\sigma_k^2 = \dfrac{dr^2}{1 - k\,r^2}+ 
r^2\,(d\vartheta^2+\sin^2\vartheta d\varphi^2) 
\end{equation}
is the metric for a spherically symmetric space with curvature 
index $k=+1,0,-1$. In a common approach our universe is 
identified with a hypersurface $\Sigma_{y_0} : y=y_0=$ constant
which is orthogonal to the extra dimension. On the hypersurface 
$\Sigma_{y_0}$, introducing  the proper time T by means of $T 
= \displaystyle{\int} n(t,y_0)\,dt$,  the line element (1) can be
reduced to the usual form of the FRW metric
\begin{equation}
ds_4^2 = a^2(T,y_0)\,d\sigma_k^2 -dT^2
\end{equation}
However this embedding is  not unique.  Ponce de Leon assumed 
that our universe is  generated on a moving hypersurface 
$\Sigma_f : y=f(t)$ which is not orthogonal to the extra  
dimension because its normal vector 
\begin{equation}
\nu_A = \dfrac{\epsilon\,n\,b}{\sqrt{n^2-\epsilon\,b^2 
\left(\dfrac{df}{dt}\right)^2}}\,\left(0,0,0,-\,
\dfrac{df}{dt},1\right)
\end{equation}
is not tangent to the $y$-lines. The metric induced on $\Sigma_f$
is
\begin{equation}
ds_4^2=a^2(t,f(t))\,d\sigma_k^2 - \left[n^2(t,f(t))-\epsilon\, 
b^2(t,f(t))\,\left(\dfrac{df(t)}{dt}\right)^2\right]\,dt^2 
\end{equation} 
The function $f(t)$ which determines the hypersurface $\Sigma_f$ 
is the solution to
\begin{equation} 
n^2(t,f(t))-\epsilon\ 
b^2(t,f(t))\,\left(\dfrac{df(t)}{dt}\right)^2 = 1 
\end{equation}
Now the scale factor depends  on $f(t)$ so the effective  matter 
content on $\Sigma_f$ is not the same as in ``conventional'' FRW 
models: the evolution of the extra dimension carries  higher 
dimensional modifications to four-dimensional general relativity.
In this paper, utilizing the  geometric construction  performed 
by Sehara and Wesson ${}^{4-6}$ to obtain a covariant foliation 
of a five-dimensiona manifold $M$, we  show that such 
modifications are obtained also in  four-dimensional 
hypersurfaces $\Sigma_\ell$ which are constant and orthogonal to 
a scalar function $\ell(t,y)$ which takes the role of the extra 
dimension in a suitable coordinate system. Let us consider  the 
five-dimensional manifold $M$ given by Eq. (1) when the  extra 
dimension $y =f(t)$ is time-dependent. The geometric construction
of Refs. 4-6, which here we briefly recall, introduces a scalar 
function $\ell=\ell(t,y)$ which defines a foliation of the 
higher-dimensional manifold $M$ with hypersurfaces $\Sigma_\ell$ 
given by $\ell =$ constant. Each hypersurface  $\Sigma_\ell$ 
corresponds to a four-dimensional spacetime and is  assumed to 
have a normal vector given by 
\begin{equation}
 n_A = \varepsilon\,\Phi\,\dfrac{\partial\ell}{\partial x^A} \,, 
\quad n_A n^A = \varepsilon 
\end{equation} 
The scalar $\Phi$  which normalizes $n^A$ is known as the lapse 
function. The projector tensor  $h_{AB}$ from the bulk to the 
hypersurfaces is 
\begin{equation}
 h_{AB}=g_{AB} - \varepsilon\, n_A n_B 
\end{equation}
This tensor is symmetric and orthogonal to
$n^A$. Each hypersurface $\Sigma_\ell$ is mapped by a 4D 
coordinate system $\{\tilde{x}^\alpha\}$. The four basis vectors 
\begin{equation}
e^A_\alpha = \dfrac{\partial  x^A}{\partial\tilde{x}^\alpha}\,  
\quad \mathrm{with}  \quad n_A e^A_\alpha = 0 
\end{equation}
are tangent to the $\Sigma_\ell$ hypersurfaces and orthogonal to 
$n_A$. These basis vectors can be used to project 5D objects 
onto $\Sigma_\ell$ hypersurfaces.  The induced metric on the 
$\Sigma_\ell$ hypersurfaces is given by
\begin{equation}
h_{\alpha\beta} = e^A_\alpha e^B_\beta g_{AB} =  e^A_\alpha 
e^B_\beta h_{AB}
\end{equation}
Clearly $\{\tilde{x}^\alpha,\ell\}$ defines an  alternative  
coordinate system to $\{x^\alpha,y\}$ on $M$. Moreover 5D vectors
are decomposed into the sum of a part tangent  to $\Sigma_\ell$ 
and a part normal to $\Sigma_\ell$. For $dx^A$ it results 
\begin{equation}
dx^A = e^A_\alpha d\tilde{x}^\alpha + \left(N^\alpha 
e^A_\alpha +\Phi\, n^A \right) d\ell 
\end{equation}
The 4D vector $N^\alpha$ is called the  shift vector and it 
describes how the $\{\tilde{x}^\alpha\}$ coordinate system 
changes as one moves from a given  $\Sigma_\ell$ hypersurface to 
another. The 5D line element (1) can then be rewritten as
\begin{widetext}
\begin{equation} 
ds_5^2 = h_{\alpha\beta}\left(d\tilde{x}^\alpha+N^\alpha 
d\ell\right)\left(d\tilde{x}^\beta+N^\beta d\ell\right) 
+\varepsilon \,\Phi^2\,d\ell^2 
\end{equation}
\end{widetext}
We choose on $M$ the coordinate system $\tilde{x}^A 
=\{\tilde{x}^\alpha,\ell\}$ alternative to ${x}^A=\{x^\alpha,y\}$
maintaining unchanged the  spatial coordinates 
$r,\vartheta,\varphi$ but changing the time coordinate from 
$t$ to $\tau$, namely $\tilde{x}^A = (r, \vartheta, \varphi, 
\tau, \ell)$, so we have to consider the following transformation
in the $(t,y)$ hyperplane: 
\begin{equation}
t=t(\tau,\ell), \quad y=y(\tau,\ell) 
\end{equation}
Then after obtaining from the diffeomorphism (11) the foliation 
parameters $\Phi$ and $N^\alpha$  the line element (12) becomes
\begin{widetext}
\begin{eqnarray}\hspace{-.3cm} 
ds_5^2 = a^2\, d\sigma_k^2 \hspace{-0.3cm}
&&-\,\left[n^2\left(\dfrac{\partial 
t}{\partial\tau}\right)^2-\, \epsilon\,b^2 
\left(\dfrac{\partial y}{\partial\tau}\right)^2\right] \left[ 
d\tau+ \dfrac{n^2\left(\dfrac{\partial t}{\partial\ell}\right) 
\left(\dfrac{\partial \ell}{\partial y}\right) +\, 
\epsilon\,b^2 \left(\dfrac{\partial y}{\partial\ell}\right) 
\left(\dfrac{\partial \ell} {\partial t}\right)} 
{n^2\left(\dfrac{\partial t}{\partial \tau}\right)\left(\dfrac 
{\partial \ell}{\partial y}\right)+\,\epsilon \, b^2 
\left(\dfrac{\partial y}{\partial \tau} \right)\left(\dfrac 
{\partial \ell}{\partial t}\right)} d\ell\right]^2 \nonumber \\
&&+\,\epsilon\,n^2 b^2\,\dfrac{\left(\dfrac{\partial t}
{\partial \ell}\right) \left(\dfrac{\partial \ell}{\partial 
t}\right) +\, \left(\dfrac{\partial y}{\partial \ell}\right)
\left(\dfrac{\partial \ell}{\partial y}\right)}
{n^2\left(\dfrac{\partial \ell} {\partial y}\right)^2-\,
\epsilon\,b^2 \left(\dfrac{\partial \ell} {\partial
t}\right)^2}\,d\ell^2 
\end{eqnarray} 
\end{widetext}
Here the functions $a$, $n$ and $b$ depend on  $\tau$ and $\ell$.
Let us notice that the line element (14) can be reduced on a 
hypersurface $\Sigma_ {\ell_0} : \ell=\ell_0$ = constant to the 
usual form of the FRW metric requiring the condition 
\begin{equation} 
\left[n^2(t(\tau),y(\tau))\left(\dfrac{d t(\tau)}{d 
\tau}\right)^2-\, \epsilon\,b^2(t(\tau),y(\tau)) \left(\dfrac{d 
y(\tau)}{d \tau} \right)^2 \right]=1
\end{equation}
which in our approach takes the place of Eq. (6) and shows that
the functions $t(\tau)$ and  $y(\tau)$ are  not independent but, 
as already found in a similar  context in Ref. 3, they can be 
parametrized by one function $F(\tau)$ of the proper time $\tau$.
The condition $n_A n^A=\epsilon$ gives 
\begin{equation}
\Phi^2=\dfrac{n^2\,b^2}{n^2\,\left(\dfrac{\partial \ell}{\partial
y} \right)^2 -\,\epsilon \,b^2\,\left(\dfrac{\partial 
\ell}{\partial t} \right)^2}
\end{equation} 
so looking at the expression of $\Phi^2$ in Eq. (14) it must be
\begin{equation}
\left(\dfrac{\partial t} {\partial \ell}\right)\, 
\left(\dfrac{\partial \ell}{\partial t}\right) + 
\left(\dfrac{\partial y}{\partial \ell}\right)\, 
\left(\dfrac{\partial \ell}{\partial y}\right) = 1 
\end{equation}
Moreover the condition $e_\alpha^A n_A=0$ gives 
\begin{equation}
\left(\dfrac{\partial t}{\partial \tau}\right)\,\left(\dfrac{
\partial \ell}{\partial t}\right) + \left(\dfrac{\partial 
y}{\partial \tau}\right)\, \left(\dfrac{\partial \ell}{\partial 
y}\right)=0 
\end{equation}
To satisfy the constraints (17) and (18) we begin choosing, 
between all the possible transformations, the following one 
\begin{equation}
\begin{cases}
t =  F(\tau)\,\cosh{(\sqrt{\epsilon} \,\psi)} 
+(\ell-\ell_0)\,\sqrt{\epsilon}\sinh{(\sqrt{\epsilon}\,\psi)} \\
y-y_0 =  F(\tau)\,\dfrac{\sinh{(\sqrt{\epsilon} \,\psi)}}{\sqrt{
\epsilon}}+(\ell-\ell_0)\,\cosh{(\sqrt{\epsilon}\,\psi)}
\end{cases}
\end{equation}
where $\psi$ is a constant. One can verify that the constraint (18) 
is satisfied solving the partial differential equation for the 
function $\ell =\ell(t,y)$ which, in a simple form, is
\begin{equation}
\ell=\ell_0-\,\dfrac{\sinh{(\sqrt{\epsilon} \,\psi)}} 
{\sqrt{\epsilon}}\,t+\cosh{(\sqrt{\epsilon} \,\psi)}\,(y-y_0)
\end{equation}
while the constraint (17) becomes an identity after substituting 
in it the derivatives of $\ell(t,y)$. Eq. (20) shows that $\ell$ 
= constant implies that here $y(t)$ is a linear function of $t$. 
The value of $F(\tau)$ can be determined once are known the 
metric coefficients $n$ and $b$.  Finally we can write the higher
dimensional line element as 
\begin{widetext}
\begin{eqnarray} 
ds_5^2=a^2\,d\sigma_k^2 
-\,\left(n^2\,\cosh^2{(\sqrt{\epsilon} 
\,\psi)}-\,b^2\,\sinh^2{(\sqrt{\epsilon} \,\psi)}\right)\Bigg[ 
\left(\dfrac{dF}{d\tau}\right)\,d\tau  \nonumber \\
\hspace{-.5cm}-\,\dfrac{\left(n^2-b^2\right)\,\sqrt{\epsilon} 
\sinh{(\sqrt{\epsilon}\,\psi)}\cosh{(\sqrt{\epsilon} \,\psi)}} 
{\left(n^2\,\cosh^2{(\sqrt{\epsilon} \,\psi)}-\,b^2\,\sinh^2{ 
(\sqrt{\epsilon} \,\psi)}\right)}\,d\ell\Bigg]^2+\epsilon\, 
\dfrac{n^2\,b^2}{\left(n^2\,\cosh^2{(\sqrt{\epsilon} 
\,\psi)}-\,b^2\,\sinh^2{(\sqrt{\epsilon}\,\psi)}\right)}\,d\ell^2
\end{eqnarray} 
\end{widetext}
We notice that while on $\Sigma_f$  it was $dy/dt\neq  0$ now on 
$\Sigma_\ell$ it results  $d\ell/d\tau =0$, so the normal 
vector  to $\Sigma_\ell$ is tangent to the $\ell$-lines.  The metric 
induced on the hypersurface $\Sigma_ {\ell_0}$ is 
\begin{equation}
ds_4^2 = a^2\,d\sigma_k^2-\,\left(n^2\,\cosh^2{(\sqrt{\epsilon} 
\,\psi)}-\,b^2\,\sinh^2{(\sqrt{\epsilon} \,\psi)}\right)\,  
\left(\dfrac{dF}{d\tau}\right)^2\,d\tau^2
\end{equation} 
Clearly on $\Sigma_{\ell_0}$ we have that $a,n$ and $b$ depend 
only on $\tau$ through the function $F(\tau)$. The line  element 
(22) can be reduced to the usual form of the FRW metric requiring
that 
\begin{equation} 
\left[n^2(F)\,\cosh^2{(\sqrt{\epsilon} 
\,\psi)}-\,b^2(F)\,\sinh^2{(\sqrt{\epsilon} \,\psi)}\right]\,  
\left(\dfrac{dF}{d\tau}\right)^2 =1
\end{equation}
which, with the initial condition $F(\tau)\arrowvert_{\tau=0}=0$, 
provides the  unknown function  $F(\tau)$. We notice that the 
left-hand side of (23) is clearly greater than zero  when 
$\epsilon =-1$ but when $\epsilon =1$ one has to discuss the sign
of the term enclosed within square brackets. The FRW  metric (3) 
is recovered from (22) in the particular case $\psi=0$ because, 
as can be checked using Eqs. (19) and (23), in this case it 
results $y=y_0$ and $T=\displaystyle{\int n(F)dF}=\tau$. 
\section{A \lowercase{comparison  between induced  metrics on 
$\boldsymbol{\Sigma_{y_0}}$, $\boldsymbol{\Sigma_{\ell_0}}$ and 
$\boldsymbol{\Sigma_{f}}$}} \pni 
To see more in detail how our    model works and to make the 
comparison between the induced metrics on the various 
hypersurfaces above defined, we shall consider the well-known 
five-dimensional metric found by Ponce de Leon ${}^7$ 
\begin{equation} 
ds_5^2 =A^2\, \left(\dfrac{t}{L}\right)^{2/\alpha} 
\,\left(\dfrac{y}{L}\right)^{2/(1-\alpha)}\,d\sigma_0^2-\, 
\left(\dfrac{y}{L}\right)^2\,dt^2+\left(\dfrac{\alpha}{1-\, 
\alpha}\right)^2\, \left(\dfrac{t}{L}\right)^2\,dy^2 
\end{equation} 
where $A$ and $L$ are constant lengths and 
$\alpha$ is a constant dimensionless parameter different from $0$
and $1$. This metric is a solution to the five-dimensional 
Einstein equations in vacuum, it is flat ($k=0$) in ordinary 
three-space and has a space-like ($\epsilon=+1$) extra dimension.
Equation (24) is one of the classes of solutions  obtained in 
Ref. 7 for cosmological models in a Kaluza-Klein theory; it was 
worked out by Wesson ${}^8$ to discuss the details of a FRW model
and it was generalized by Rippl, Romero and Tavakol ${}^9$ to 
study lower-dimensional gravity. Since then,  many  other 
cosmological solutions and their associated matter properties 
have been derived and the whole analysis about the  embedding 
of four-dimensional general relativity in five dimension goes 
back to the Campbell theorem which was rediscovered by Romero, 
Tavakol and Zalaletdinov. ${}^{10}$ First we recall some results 
obtained projecting the metric (24) on the hypersurface 
$\Sigma_{y_0}$ where the line element is 
\begin{equation} 
ds_4^2=A^2\,\left(\dfrac{y_0}{L}\right)^{2/(1-\alpha)}\, 
\left(\dfrac{T}{y_0}\right)^{2/\alpha}\,d\sigma_0^2 - dT^2 
\end{equation} here $T=(y_0/L)\,t$ is the FRW proper time. 
Consequently the pressure $p$, the density $\rho$,  the equation 
of state $p/ \rho=w$ of the induced matter, the gravitational 
density $\rho_ g=3\,p+\rho$ and the deceleration parameter $q$ 
are given by \begin{equation} 8\pi p=\dfrac{2\alpha -3}{\alpha^2 
T^2} \, , \quad 8\pi \rho=\frac{3}{\alpha^2 T^2} \, ,
\quad \dfrac{p}{\rho}=\left(\dfrac{2\alpha -3}{3}\right) \, ,
\quad \rho_g= \dfrac{3\,(\alpha-1)}{4\pi \alpha^2 T^2} \, ,
\quad q=\alpha -1 
\end{equation} 
So models with $\alpha \in(0,1)$ describe an accelerating 
universe with exotic matter, while models with $\alpha >1$ have 
ordinary matter satisfying the strong energy condition. Models 
with $\alpha < 0$ are excluded because they imply a contracting 
universe. The present-day age $T_0$  of an universe emerging from
a big bang  is given, as explicitly first pointed out in Ref. 
[11], by 
\begin{equation}
T_0=\dfrac{H_0^{-1}}{1+\bar{q}(T_0)}
\end{equation}
where $H_0$ and $\bar{q}(T_0)$ are respectively  the present-day 
values of the Hubble parameter and of the average  deceleration 
parameter which is given by
\begin{equation}
\bar{q}(T_0)=\dfrac{1}{T_0}\displaystyle{\int_0^{T_0}}q(T)dT
\end{equation} 
If the conjecture $H_0T_0=1$ is valid then the average 
deceleration parameter must be zero when averaged after a long 
interval of time, which means that the universe evolves through 
a cascade of accelerating/decelerating regimes. Our aim in this 
paper is to show that higher-dimensional modifications to FRW 
spacetimes can be obtained also on 4D static hypersurface,  so we
shall not treat here the cross-over from decelerate to accelerate
cosmic expansion. Now we give our results obtained projecting the
metric (24) on the hypersurface $\Sigma_{\ell_0}$. Equation (23) 
for the function $F(\tau)$ becomes 
\begin{equation} 
\dfrac{\cosh^2{(\psi)}}{L^2}\,\left[ 
\dfrac{(1-2\,\alpha)}{(1-\alpha)^2}\,\sinh^2{(\psi)}\,F^2(\tau) 
+2\,y_0\,\sinh{(\psi)}\,F(\tau)+y_0^2\right]\, 
\left(\dfrac{dF(\tau)}{d\tau}\right)^2=1 
\end{equation} 
Before solving Eq. (29) we have to discuss the sign of the 
binomial in $F(\tau)$ enclosed within square brackets. Hereafter 
we shall assume that $\psi$, $y_0$ and $L$ are all positive and 
finite constants and we shall require that $F(\tau)$ increases 
with $\tau$. The binomial is greater than zero: i) if 
$0<\alpha \leq 1/2$ for all $F(\tau)>0$; ii) if $1/2<\alpha <1$ 
for $F(\tau)<(y_0/\sinh{(\psi)})\,(1-\alpha)/ (2\alpha-1)$; iii) 
if $\alpha >1$ for $F(\tau)<(y_0/\sinh{(\psi)}) \,(\alpha -1)$. 
Once $F(\tau)$ has  been found the scale factor for the metric 
(24) becomes 
\begin{equation} 
a(\tau)=A\,\left(\dfrac{F(\tau)}{L}\cosh{(\psi)}\right)^{1/\alpha}
\left(\dfrac{y_0}{L}+ \dfrac{F(\tau)}{L}\sinh{(\psi)}\right)^{1/
(1-\alpha)}
\end{equation}
Now the scale factor depends  on $F(\tau)$ so the effective 
matter content on $\Sigma_{\ell_0}$ is not the same as in 
``conventional'' FRW models. It is apparent that when 
$\alpha>1/2$ the scale factor can not become greater than a 
particular amount in time, so models with $\alpha>1/2$ can 
describe a particular stage of the evolution of the universe, for
example its early evolution. Two consecutive stages can be joined
by using appropriate junction conditions.${}^{1,2}$ Equation 
(29) can be easily integrated, however in the case $\alpha \neq 
1/2$ it is not possible to explicitly obtain $F(\tau)$ as a 
function of $\tau$ so one must use an approximate expression for 
it. Let us begin considering the value $\alpha=1/2$ for which the
function $F(\tau)$ can be exactly obtained. It is worth noticing 
that Seahra and Wesson ${}^{12}$ discussing the structure of the 
big bang from higher-dimensional embeddings mentioned that the 
$\alpha=1/2$ cosmology is the only case for which the Ponce de 
Leon metric (24) is well defined. From Eq. (29) we have 
\begin{equation}
F(\tau)=\dfrac{y_0}{2\, 
\sinh{(\psi)}}\,\left[(1+\kappa\, \tau)^{2/3}-1\right]
\end{equation}
where $\kappa=(3 L/y_0^2)\,\tanh{(\psi)}$. The scale factor is
\begin{equation}
a(\tau)=A\,\left[\dfrac{1}{4 \tanh{(\psi)}}\left(\dfrac{y_0}{L}  
\right)^2\right]^2\,\left[(1+\kappa\, \tau)^{4/3}-1\right]^2
\end{equation} 
Pressure and density are
\begin{equation}
8\pi p=-\,\dfrac{16}{9}\kappa^2\,\dfrac{\left[9\,(1+\kappa\, 
\tau)^{4/3}-1\right]}{(1+\kappa\,\tau)^{2/3}\,\left[(1+\kappa\, 
\tau)^{4/3}-1\right]^2} \, ,  \quad
8\pi \rho= \dfrac{64}{3}\kappa^2\,\dfrac{(1+\kappa\,
\tau)^{2/3}}{\left[(1+\kappa\, \tau)^{4/3}-1\right]^2} 
\end{equation}
and the equation of state of the effective matter is
\begin{equation}
\dfrac{p}{\rho}=-\dfrac{3}{4}\,\left(1-\dfrac{1}{9\,(1+\kappa\, 
\tau)^{4/3}}\right)
\end{equation}
Gravitational density and deceleration parameter are
\begin{equation}
\rho_g=-\,\dfrac{2}{3\pi}\, \dfrac{\left[5\,(1+\kappa\, 
\tau)^{4/3}-1\right]}{(1+\kappa\,\tau)^{2/3}\,\left[(1+\kappa\, 
\tau)^{4/3}-1\right]^2} \, ,  \quad
q=-\,\dfrac{5}{8}\,\left(1-\dfrac{1}{5\,(1+\kappa\, 
\tau)^{4/3}}\right)
\end{equation} 
In the case $\alpha=1/2$ we have therefore an accelerating 
universe with exotic matter. Before making comparison  with  
observers in the  hypersurfaces $\Sigma_{y_0}$ and $\Sigma_f$ 
we recall that they use different clocks and that the relations 
between the proper times $T,t$ and $\tau$ are given by 
\begin{equation}
T=\dfrac{y_0}{L}\,t= \dfrac{y_0}{L}\,F(\tau)\,\cosh{(\psi)}
\end{equation}
From  the results found  on  $\Sigma_{y_0}$ it is  apparent that 
there are higher-dimensional modifications to the FRW spacetime 
(25). The usual FRW description is however  approximately 
recovered in a period close enough to the initial time $\tau=0$. 
Starting again from the  metric (24) modifications to the 
FRW spacetime (25) were first  obtained by Ponce de Leon.${}^{1,2}$ 
In particular on the hypersurface $\Sigma_{f}$ when 
for $\alpha =1/2$ it results 
\begin{equation}
f(t)= \dfrac{L}{2\,K\,t}\,\left(1+K^2\,t^2\right) 
\end{equation}
\begin{equation}
a(t)= \dfrac{A}{(2 C)^2}\,\left(1+K^2\,t^2\right)^2
\end{equation}
\begin{equation}
8\pi p=-\,\dfrac{8\,K^2\,\left(1+5\,K^2\,t^2\right)} 
{\left(1+K^2\,t^2\right)^2}, \quad 
8\pi \rho = \dfrac{48\,K^4\,t^2}{\left(1+K^2\,t^2\right)^2}, 
\quad \dfrac{p}{\rho}=-\dfrac{5}{6}\,\left(1+\dfrac{1} 
{5\,K^2\,t^2}\right) 
\end{equation}
\begin{equation}
\rho_g=-\dfrac{3}{\pi}\,\dfrac{\left(1+3\,K^2\,t^2\right)}
{\left(1+K^2\,t^2\right)},  \quad
q=-\dfrac{3}{4}\,\left(1+\dfrac{1}{3\,K^2\,t^2}\right)
\end{equation}
where $C$ is a dimensionless constant coming from the 
integration of Eq. (6) and  $K=C/L$.  We have again an 
accelerating universe with  exotic matter but the  
higher-dimensional modifications are different from those found 
on $\Sigma_{\ell_0}$. Equation (38) gives $a(\tau)
\arrowvert_{\tau=0}\neq0$ in disagreement with FRW  models, 
however this feature can be put away ${}^{1,2}$  making use of 
the dominant energy condition which requires  $Kt\geq1$. When
$\alpha \neq 1/2$ we integrate Eq. (29) but now  $F(\tau)$ cannot
be explicitly given as a function of $\tau$ to obtain  $a(\tau)$ 
on $\Sigma_{\ell_0}$. We shall therefore consider approximate 
expressions for $F(\tau)$ which can be obtained from (29) using 
power series expansions in the particular cases when $F(\tau)/L 
\ll 1$  (early evolution of the universe) and when $F(\tau)/L \gg
1$ (late evolution of the universe). After the values of 
$F(\tau)$ have been found we shall obtain the corresponding 
approximate values of the scale factor $a(\tau)$ and of the other
quantities of interest by a series expansion in the proper time 
both in the early universe $\kappa\tau \ll 1$ and in the late 
universe $\kappa\tau\gg 1$. Finally, the expressions of 
$F(\tau),\, a,\, p,\,\rho, \,p/\rho, \,\rho_g$, $q$ are given in 
Appendix A when $\kappa\tau \ll 1$ and in Appendix B when 
$\kappa\tau\gg 1$. Since the quantities of physical interest 
can be derived from the knowledge of the scale factor,   to 
compare between our results and those obtained on 
$\Sigma_{y_0}$ and $\Sigma_{f}$ we recall that the expression of 
the scale factor on $\Sigma_{y_0}$ is 
\begin{equation} 
a(T)= A\,\left(\dfrac{y_0}{L}\right)^{1/(1- \alpha)}\, 
\left(\dfrac{T}{y_0} \right)^{1/\alpha} 
\end{equation} 
and on 
$\Sigma_{f}$ is  
\begin{equation} 
a(t)=\dfrac{A}{(2C)^{1/\alpha}}\left(\dfrac{t}{L}\right)^{\left( 
1-|1-\alpha|/(1-\alpha)\right)/\alpha}\left[1+C^2\left(\dfrac{t} 
{L}\right)^{2|1-\alpha|/\alpha}\right]^{1/(1-\alpha)}  
\end{equation}
It is apparent from Appendix A that the usual FRW description is 
approximately recovered in a period close enough to the 
initial time $\tau=0$. Later on, both Appendixes show that  
when $\alpha\neq1/2$ there are on $\Sigma_{\ell_0}$ 
higher-dimensional modifications to the ``conventional'' FRW 
spacetime (25) and that these modifications are different from 
the ones found on $\Sigma_f$. As emphasized in Refs. 1-2 the 
different results coming from different values of $\alpha$ 
represent the same spacetime in another parametrization so if the
value of $\alpha$ is allowed to change in order to have a not 
constrained equation of state, then one can study a more 
realistic model of the universe by joining metrics with different
values of $\alpha$ across a time-constant hypersurface. Such a 
calculation is, however, beyond the scope of this paper. 
\section{C\lowercase{onclusion}} 
We have shown that, starting from a 5D   given metric which has a
large time-dependent extra dimension and describes a homogeneous 
universe, it is possible to consider an embedded static 
hypersuface  $\Sigma_{\ell_0}$ where exist higher-dimensional 
modifications to the energy-momentum tensor as obtained on a 
static FRW hypersurface $\Sigma_{y_0}$ and also different from 
those found  on a dynamic hypersurface $\Sigma_f$. As a working 
example, we applied our model to the line element (24) which can 
be considered the generalization of the flat FRW cosmological 
metric to five dimensions so we have  to check whether the 
obtained results are in agreement with the present-day 
observational data, in particular with the fact that our universe
is now in accelerated expansion. Irrespective of the value 
assumed by the quantity $\kappa \,\tau$ at the present age 
$\tau_0$ of the universe, we see that when $\alpha <1$ it results
$q<0$ but if we limit the range of $\alpha$ to $\alpha \leq 1/2$,
then we have that $q \in (-1,-1/2)$ which is consistent with 
recent constraints on accelerating universe combined with various
cosmological probes. ${}^{13}$ It remains to fix the value of 
$\kappa\,\tau$ today, but having choosen to associate 
$\kappa\,\tau \ll 1$ and $\kappa\,\tau \gg 1$ respectively to the
early and to the late universe, we choose $\kappa\,\tau_0  = 1$ 
as an acceptable value. Therefore we must use  the exact 
solution corresponding to $\alpha = 1/2$ and find that today the 
deceleration parameter is $q_0 = -0.575$ while the equation of 
state parameter is $w_0 =-0.717$, results which are in accordance
with the estimated mean values of these parameters in the actual 
universe. Finally, in the hypotheses that the free parameter 
$\alpha$ in the considered line element (24) can be a function of
the time $\tau$ and that the values $\kappa\,\tau \ll 1$ and 
$\kappa\,\tau \gg 1$ correspond respectively to the early 
(post-inflationary) and to the late  universe, one might 
associate different values of $\alpha$ to the various eras of the
evolving universe. So, for example, when $1< \alpha <3$ the 
primordial matter behaves similar to ordinary matter and the 
expansion is slowing down, but when $0< \alpha <1$ the matter has
exotic properties, and the expansion is speeding up. It will be 
interesting to apply the approach we suggest in this paper to 
other solutions of modern cosmologies with extra dimensions. 
\appendix 
\section{R\lowercase{esults on} $\boldsymbol{\Sigma_{\ell_0}}$ 
\lowercase{when} $\boldsymbol{\alpha > 1/2}$} 
When $\alpha >1/2$ we obtain in the early universe  
($\kappa\tau \ll 1$) 
\begin{equation} 
F(\tau)\approx\dfrac{y_0}{\sinh{(\psi)}}\,\left[\sqrt{1+ 
\dfrac{2}{3}\kappa\tau}-1\right] 
\end{equation}
\begin{equation}
a(\tau)\approx A\,\dfrac{1}{\left(3\tanh(\Psi)\right)^{1/
\alpha}}\left(\dfrac{{y_0}}{L}\right)^{ 1/(\alpha (1-
\alpha))}\left(\kappa\,\tau\right)^ {1/\alpha}\left[1-
\dfrac{3\alpha -1}{6\alpha (\alpha -1)} \kappa\tau\right] 
\end{equation} 
\begin{equation}
8\pi p\approx\dfrac{2\alpha -3}{\alpha^2 \tau^2}\left[1+
\dfrac{3\alpha -1}{(2\alpha -3)(\alpha -1)}\kappa\tau\right],
\quad
8\pi\rho\approx \dfrac{3}{\alpha^2 \tau^2}\left[1-\dfrac
{3\alpha-1}{3(\alpha-1)}\kappa\tau\right]
\end{equation}
\begin{equation}
\dfrac{p}{\rho}\approx\dfrac{2\alpha -3}{3}\left[
1+\dfrac{2\alpha (3\alpha -1)}{3(2\alpha -3)(\alpha -1)}
\kappa\tau\right],
\quad
\rho_g\approx\frac{3(\alpha-1)}{4\pi\alpha^2\tau^2}\left[
1+\dfrac{3\alpha -1}{3(\alpha -1)^2}\kappa\tau\right]
\end{equation}
\begin{equation}
q\approx (\alpha-1)\left[1+\dfrac{\alpha(3\alpha -1)}{3(\alpha 
-1)^2}\kappa\tau\right]
\end{equation}
\section{R\lowercase{esults on} $\boldsymbol{\Sigma_{\ell_0}}$ 
\lowercase{when} $\boldsymbol{\alpha < 1/2}$}  
When $\alpha<1/2$  we obtain in the late universe 
($\kappa\tau \gg 1$) 
\begin{equation}
F(\tau)\approx\dfrac{y_0}{\sinh{(\psi)}}\dfrac{(1-\alpha)^2}
{(1-2\alpha)}\left[\sqrt{1+\dfrac{2(1-2\alpha)^{3/2}}{3(1-
\alpha)^3}\kappa\tau}-1\right]
\end{equation} 
\begin{widetext} 
\begin{eqnarray}
a(\tau)\approx 
A\,\left[\dfrac{(1-\alpha)}{\tanh^{2(1- 
\alpha)}{(\Psi)}\,\sqrt{1-2\alpha}}\left(\dfrac{y_0}{L} 
\right)^2 \right]^{1/{(2 \alpha (1-\alpha))}} 
\left(\kappa\,\tau\right)^{1/{(2 \alpha (1-\alpha))}}  
\nonumber \\ 
\left[1-\dfrac{\sqrt{3}(1-3\alpha(1-\alpha))}{\sqrt{2}
\alpha\left((1-\alpha)\,\sqrt{1-2\alpha}\right)^{3/2}}
\,\dfrac{1} {\sqrt{\kappa\tau}}\right] 
\end{eqnarray}
\end{widetext} 
\begin{equation}
8\pi p\approx -\,\dfrac{(3-4\alpha(1-\alpha))}{4\alpha^2
(1-\alpha )^2\,\tau^2} \,\left[1-\dfrac{3\sqrt{6} (1-\alpha
(1-\alpha))(1-3\alpha(1-\alpha))}{(3-4\alpha(1-\alpha) 
)\sqrt{(1-\alpha)\,(1-2\alpha)^{3/2}}}\,\dfrac{1}
{\sqrt{\kappa\tau}}\right]
\end{equation}
\begin{equation}
8\pi \rho \approx \dfrac{3}{4\alpha^2(1-\alpha )^2\,\tau^2}\,
\left[1+\dfrac{\sqrt{6}(1-3\alpha(1-\alpha))}{\sqrt{(1-\alpha)
\,(1-2\alpha)^{3/2}}}\,\dfrac{1}{\sqrt{\kappa\tau}}\right] 
\end{equation}
\begin{equation}
\dfrac{p}{\rho}\approx -\,\left(\dfrac{3-4\alpha(1-\alpha)}
{3}\right)\, \left[1+\dfrac{\sqrt{6}\alpha(1-3\alpha(1-
\alpha))}{(3-4\alpha(1-\alpha))}\sqrt{\dfrac{(1-\alpha)}{(1
-2\alpha)^{3/2}}}\,\dfrac{1} {\sqrt{\kappa\tau}}\right] 
\end{equation} 
\begin{equation}
\rho_g\approx -\,\dfrac{3(\alpha^2+(1-\alpha)^2)}{16\pi 
\alpha^2(1-\alpha^2)\,\tau^2}\,\left[1-\dfrac{\sqrt{3}(1-3\alpha(1-\alpha))(2- 
3\alpha(1-\alpha))}{\sqrt{2}(1-2\alpha(1-\alpha))\sqrt{ 
(1-\alpha) \,(1-2\alpha)^{3/2}}}\,\dfrac{1}{\sqrt{\kappa 
\tau}}\right] 
\end{equation} 
\begin{equation}
q \approx 
-\,(1-2\alpha(1-\alpha))\,\left[1+\dfrac{\sqrt{3}\alpha(1-3\alpha
(1-\alpha))}{\sqrt{2}(1-2\alpha(1-\alpha))}\sqrt{\dfrac{(1-\alpha)}
{(1-2\alpha)^{3/2}}}\,\dfrac{1} {\sqrt{\kappa\tau}} \right] 
\end{equation}
 
\end{document}